\documentclass[10pt,letterpaper]{article}
\usepackage[top=0.85in,left=2.75in,footskip=0.75in]{geometry}

\usepackage{amsmath,amssymb}
\usepackage{changepage}
\usepackage{textcomp}
\usepackage{cite}
\usepackage{nameref,hyperref}
\usepackage[right]{lineno}
\usepackage[nopatch=eqnum]{microtype}
\DisableLigatures[f]{encoding = *, family = * }
\usepackage[table]{xcolor}
\usepackage{array}
\usepackage{booktabs}

\newcolumntype{+}{!{\vrule width 2pt}}
\newlength\savedwidth

\newcommand\thickhline{\noalign{\global\savedwidth\arrayrulewidth\global\arrayrulewidth 2pt}%
\hline
\noalign{\global\arrayrulewidth\savedwidth}}

\raggedright
\usepackage[aboveskip=1pt,labelfont=bf,labelsep=period,justification=raggedright,singlelinecheck=off]{caption}

\makeatletter
\renewcommand{\@biblabel}[1]{\quad#1.}
\makeatother

\usepackage{lastpage,fancyhdr,graphicx}
\usepackage{epstopdf}
\fancyheadoffset[L]{2.25in}
\fancyfootoffset[L]{2.25in}
\begin{document}
\vspace*{0.2in}

\begin{flushleft}
{\Large
\textbf{Do stationarity transformations improve time series forecasts? A controlled experimental evaluation}
}
\newline
\\
Bhanu Suraj Malla\textsuperscript{1*},
Yuqing Hu\textsuperscript{1}
\\
\bigskip
\textbf{1} School of Computer Science, Georgia Institute of Technology, Atlanta, GA 30332, United States of America
\\
\bigskip

* bmalla7@gatech.edu

\end{flushleft}

\section*{Abstract}
Stationarity transformations, such as differencing, are a common preprocessing step in forecasting, motivated by the idea that modifying a series to achieve stationarity improves accuracy. Whether this is true, and for which processes, has rarely been evaluated in controlled experiments. We study the decision to transform as the object of inquiry. We cross eighteen synthetic data-generating processes, most of them stochastic-trend processes spanning exact and near unit roots, fractional integration, seasonal unit roots, structural breaks, and heteroscedasticity, with ten transformations, five models, and three horizons, replicated by Monte Carlo, for 35,099 evaluations. Each forecast is inverted to the original scale, with the differencing inverse anchored at the forecast origin, and scored by the mean absolute scaled error. Signal-preserving transforms, namely deterministic detrending and seasonal differencing matched to series structure, improve accuracy, whereas indiscriminate differencing degrades it. A mediation analysis shows that differencing achieves trend stationarity, but trend stationarity is only weakly associated with accuracy, and transforms differ in their effects on predictable structure. Choosing the transformation by out-of-sample validation yields lower regret than unit-root pretesting or any fixed rule, with blanket differencing performing the worst. The findings are confirmed by real-world validation on nine series from two domains.

\clearpage
\newgeometry{top=0.85in,left=1in,right=1in,footskip=0.75in}
%\linenumbers % disabled for arXiv

\section*{Introduction}

In time series forecasting, it is common to apply transformations that impose stationarity on the variables. It is often recommended to difference, log, and/or seasonally adjust a series before applying a model, because AR, ARMA, or ARIMA models presume stationary inputs~\cite{ref1,ref2}. Following Box--Jenkins, one should difference the series until it is stationary; once stationarity cannot be rejected, one models the residuals.

This advice encodes a choice. For any time series, the forecaster must decide between differencing, detrending, stabilizing variance, or leaving the data untouched, and the two canonical forms of mean non-stationarity require opposite treatments. A trend-stationary (TS) process, $y_t = \mu + \delta t + u_t$ with $u_t$ stationary, has shocks that die away and is best handled by removing the deterministic trend; differencing it over-differences, inducing a non-invertible moving-average component and throwing away predictable low-frequency structure~\cite{ref3,ref4}. A difference-stationary process can be written as $\Delta y_t = \delta + u_t$ and contains a unit root, so it needs differencing, whereas detrending leaves the stochastic trend in place. As TS and DS processes are statistically close to indistinguishable in finite samples~\cite{ref5,ref6}, the choice is made under genuine uncertainty, and making the wrong choice results in an accuracy penalty rather than an efficiency penalty.

Unit-root pretesting usually resolves the decision. The augmented Dickey--Fuller test (ADF)~\cite{ref7}, which was developed by Said and Dickey~\cite{ref8}, tests the null of a unit root against the alternative of stationarity with ARMA errors. This is complemented by Phillips--Perron~\cite{ref3} and KPSS~\cite{ref9}. A recurring issue is a lack of power against near-unit-root alternatives: under the local-to-unity parameterization $\rho_n = 1 + c/n$ of Phillips~\cite{ref10,ref11}, asymptotic theory's singular point is the unit-root case, and in its neighborhood, tests cannot reliably separate TS from DS, which is precisely the area where the preprocessing decision matters most. Elliott, Rothenberg, and Stock's~\cite{ref12} point-optimal tests and Ng and Perron's~\cite{ref13} lag selection enhance the test as a classifier, but do not demonstrate that test-based classification is the best foundation for the forecasting decision.

Costs of the wrong transformation choice are understood at the level of the data-generating process. Over-differencing, according to Plosser and Schwert~\cite{ref14}, creates a non-invertible moving-average root that damages estimation and forecasting. In contrast, Nelson and Kang~\cite{ref15} illustrate the symmetric error, in which detrending a DS series creates spurious periodicity. Furthermore, Chan, Hayya, and Ord~\cite{ref16} emphasize how the right trend-removal technique varies by the trend type. Finally, Granger and Newbold~\cite{ref17} show how a troubled relationship, in which trends receive the wrong treatment, creates other spurious relationships. The work of Granger and Joyeux~\cite{ref18} and Hosking~\cite{ref19} on fractional integration, where $d$ can take on non-integer values, shows that a process can have long memory yet remain stationary for $d < 0.5$. That is, the operator $(1-B)^d$ eliminates long-range dependence while maintaining more structure than integer differencing~\cite{ref20,ref21}. It is a distinct family~\cite{ref22}, whose limiting case is the logarithm, which stabilizes variance but does not high-pass filter the conditional mean~\cite{ref23}. Proietti and L\"{u}tkepohl~\cite{ref24} find it useful for a large fraction of macroeconomic series. Thus, the value of a transformation is an empirical issue.

Diebold and Kilian~\cite{ref25} formulated the decision to transform in a decision-theoretic way. Three strategies---always difference, never difference, and pretest---are compared for the AR(1)-with-trend process. Pretesting routinely improves on differencing under stated conditions when modeling in levels. Their analysis is the nearest precedent, and it is the point from which this work deviates. Modern practice has three features that are not included. First, their characterization is for a single process class and a single model, whereas practitioners apply the decision across heterogeneous processes using models such as exponential smoothing in state-space form~\cite{ref26}, structural time-series models~\cite{ref27}, and gradient-boosted trees~\cite{ref28}, whose internal trend handling differs. Second, a scale-free measure that remains defined even as a series approaches zero, the mean absolute scaled error (MASE)~\cite{ref29}, rather than loss in squared error, is the proper measure for accuracy. In addition, a forecaster can choose the transformation through direct out-of-sample validation, a model-free alternative to pretesting. However, its performance has not been established relative to the unit-root-test rule.

The same tension has surfaced recently in deep learning. Stationarizing the input aids predictability but can remove structure that is itself informative, a phenomenon termed over-stationarization, and learned or reversible normalization schemes have been proposed to normalize the input and then restore its statistics at the output~\cite{ref40,ref41}. These methods are architecture-specific components tuned for a single family of neural models, and they treat normalization as an internal, always-on operation rather than as the forecaster's explicit decision of whether, and with which transform, to preprocess a series. Our study is the classical-model complement to this line: we treat the transformation as a discrete decision, evaluate it in a controlled design across heterogeneous processes and five model families, and show that the same core tension---stationarity achieved at the cost of predictable structure---governs the outcome.

To address this gap, we conduct a controlled experiment that places the choice of transformation as the object of study and disentangles two questions that the conventional pipeline lumps together: does a transformation render the time series stationary, and does this improve accuracy? We make four contributions: (1) a balanced benchmark containing eighteen data-generating processes (DGPs), including exact and near unit roots~\cite{ref10,ref11} and fractionally integrated series~\cite{ref18,ref19}, along with deterministic-trend, seasonal-unit-root, structural-break, and heteroscedastic processes; (2) evaluation on the original scale, where every differencing inverse maps forecasts from the transformed scale back to the original level scale, anchored at the last training value (the forecast origin), and accuracy is reported in MASE~\cite{ref29}; (3) a mediation decomposition estimated per transform family to assess whether transformations achieve stationarity, whether achieved stationarity improves accuracy, and the resulting net effect; and (4) a comparison of five preprocessing selectors by their regret relative to the per-series oracle, extending~\cite{ref25} to multiple model families and a heterogeneous process set. The findings corroborate classical theory where it applies: differencing a deterministic-trend series worsens accuracy~\cite{ref14,ref15}, and the transforms that improve on no preprocessing are the signal-preserving ones. Most importantly, picking the transformation by out-of-sample validation cuts regret roughly in half relative to any fixed rule and beats unit-root pretesting.

The paper proceeds as follows. The Materials and methods section covers the experimental design, data-generating processes, transformations, models, and evaluation framework. The Results section presents the main findings. The Mediation analysis section decomposes the stationarity-to-accuracy pathway. The Real-world validation section tests the findings on empirical data. The Discussion section interprets the results and addresses limitations.

\section*{Materials and methods}

The study uses a fully crossed factorial design over 18 DGPs, 10 transformations, 5 models, and 3 horizons, replicated by Monte Carlo: each of the $R$ independent realizations of every process is run through the entire grid. In each replication, we draw an independent realization of every process, apply every transformation, fit every model at every horizon, invert every forecast back to the original scale, and score it. For one replication, this yields $18 \times 10 \times 5 \times 3 = 2{,}700$ evaluations; cells in which a model fails to estimate are recorded as missing and are not imputed. The repository identified in the Data Availability Statement contains all code, configurations, and results.

\subsection*{Data-generating processes}

Each synthetic series has length $n = 220$ at weekly frequency with seasonal period $s = 52$. This satisfies the condition $2s < n$, which is required for seasonal differencing and seasonal-trend decomposition. A general additive series is given as
\begin{equation}
y_t = \mu_t + \tau_t + \varepsilon_t,
\end{equation}
where $\mu_t$ is a trend (deterministic, $\mu_t = \beta t$, or stochastic, via a unit root), $\tau_t = A\sin(2\pi t/s)$ is a seasonal component when present, and $\varepsilon_t$ is noise whose variance may depend on $t$. A near-unit-root process is an AR(1), $y_t = \rho y_{t-1} + \eta_t$, with $\rho \in \{0.90, 0.95, 0.99\}$, giving local-to-unity values $c = n(\rho-1)$ of about $-22$, $-11$, and $-2$. The parameters become progressively closer to the unit-root singularity, where the trend type is least identifiable~\cite{ref10,ref11}. An ARFIMA$(0, d, 0)$ process with $d \in \{0.2, 0.4\}$ is stationary because $d$ lies within the stationary range~\cite{ref19}. A seasonal-unit-root process is $y_t = y_{t-s} + \eta_t$ (Table~\ref{table1}).

\begin{table}[!ht]
\centering
\caption{\textbf{Synthetic data-generating processes.} The matched transform targets the process's known non-stationarity.}
\begin{tabular}{l l l l}
\hline
\textbf{Class} & \textbf{Process} & \textbf{Specification} & \textbf{Matched transform} \\
\thickhline
Stationary & baseline & $y_t = \varepsilon_t$ & none \\
\hline
Stationary & stationary\_ar2 & AR(1), $\varphi = 0.6$ & none \\
\hline
Trend-stationary & linear\_trend & $\mu_t = 0.5t$ & detrend \\
\hline
Trend-stationary & quadratic\_trend & $\mu_t = 0.1t + 0.002t^2$ & detrend \\
\hline
Trend-stationary & trend\_seasonal & $\mu_t = 0.5t$, $A = 10$ & detrend \\
\hline
Difference-stationary & random\_walk & $\Delta y_t = \eta_t$ & difference \\
\hline
Difference-stationary & random\_walk\_drift & $\Delta y_t = 0.3 + \eta_t$ & difference \\
\hline
Near-unit-root & near\_unit\_root\_90/95/99 & AR(1), $\rho = 0.90/0.95/0.99$ & difference / detrend \\
\hline
Fractional & arfima\_d20/d40 & ARFIMA$(0, 0.2/0.4, 0)$ & fractional difference \\
\hline
Seasonal unit root & seasonal\_unit\_root & $y_t = y_{t-s} + \eta_t$ & seasonal difference \\
\hline
Seasonal-stationary & deterministic\_seasonal & $A = 10$, fixed phase & seasonal difference \\
\hline
Structural break & level\_break & level shift at $t = n/2$ & none \\
\hline
Structural break & trend\_break & slope change at $t = n/2$ & detrend \\
\hline
Heteroscedastic & increasing\_variance & $\sigma_t$ linear in $t$ & log / Box-Cox \\
\hline
Heteroscedastic & multiplicative\_variance & $\sigma_t \propto |\mu_t|$ & log / Box-Cox \\
\hline
\end{tabular}
\label{table1}
\end{table}

The design includes eight stochastic-trend processes (two difference-stationary, three near-unit-root, two fractional, and one seasonal-unit-root), making them the largest group (8 of 18). The design does not presuppose that differencing is unhelpful; deterministic-trend processes are a minority (3 of 18).

\subsection*{Transformations}

We evaluate the untransformed series, six single transformations, seasonal-trend (STL) detrending, and three transformation pipelines. The single transformations are first differencing, seasonal differencing, fractional differencing, the log transform, the Box--Cox transform, and linear detrending. All transformations are fitted only on the training segment, and all forecasts are inverted to the original scale before scoring. Correct inversion requires anchoring to the appropriate reference value; otherwise, a systematic level error is introduced. Let the training series be $y_{1:T}$, where $T$ is the forecast origin. For first differencing, $w_t = y_t - y_{t-1}$, so the inverse is
\begin{equation}
\hat{y}_{T+k} = y_T + \sum_{j=1}^{k} \hat{w}_{T+j},
\end{equation}
anchored at $y_T$, the last training level. For seasonal differencing, $w_t = y_t - y_{t-s}$, and the inverse is
\begin{equation}
\hat{y}_{T+k} = \hat{y}_{T+k-s} + \hat{w}_{T+k},
\end{equation}
where $\hat{y}_{T+k-s}$ is known from training data if $k \leq s$ and is a reconstructed forecast otherwise. We estimate $d$ for fractional differencing using the Geweke--Porter-Hudak log-periodogram regression~\cite{ref20} and clip it to $(0.01, 0.49)$. We apply $(1-B)^d$ via its binomial expansion and invert it via the corresponding expansion of $(1-B)^{-d}$. For the log transform, we add an offset when the series is non-positive. For Box--Cox, we estimate the parameter by maximum likelihood. Cases in which the inverse is numerically undefined are recorded as failures. For linear detrending, we remove the ordinary-least-squares fit $\hat{a} + \hat{b}t$ and add the trend back at future indices; for STL detrending, we remove the seasonal-trend decomposition and extrapolate the terminal trend slope.

\subsection*{Models, horizons, and metrics}

We evaluate five forecasting models that can handle trends internally: AR (autoregressive) and ARIMA (autoregressive integrated moving average), which belong to the Box--Jenkins family; exponential smoothing (ETS) and the structural unobserved-components model (UCM), which derive from the state-space tradition~\cite{ref26,ref27,ref30,ref31}; and gradient boosting (GB), which belongs to the machine-learning family~\cite{ref28}. AR, ARIMA, ETS, and UCM are implemented in statsmodels~\cite{ref32}, whereas GB uses LightGBM~\cite{ref28}. In these implementations, the usual distinction between models that require stationary inputs and those that do not is blurred: AR is fitted with a deterministic-trend search over \{none, constant, constant-plus-linear\}, and UCM with a local-linear-trend component includes an internal stochastic trend, so both can accommodate non-stationarity that a conventional pipeline would remove. Table~\ref{table2} presents the hyperparameter grids; statistical-model parameters are selected by minimizing the in-sample one-step-ahead error.

\begin{table}[!ht]
\centering
\caption{\textbf{Forecasting models and hyperparameter search spaces.}}
\begin{tabular}{l l l}
\hline
\textbf{Model} & \textbf{Family} & \textbf{Hyperparameters} \\
\thickhline
AR & Box-Jenkins & lag $\in \{1,2,3,5\}$; trend $\in$ \{none, const, const+linear\} \\
\hline
ARIMA & Box-Jenkins & order $(2,1,2)$; maximum likelihood \\
\hline
ETS & State space & additive trend; estimated smoothing parameters \\
\hline
UCM & State space & local linear trend; maximum likelihood \\
\hline
GB & Machine learning & 100 trees; 15 leaves; depth 5; learning rate 0.1; lag window 12 \\
\hline
\end{tabular}
\label{table2}
\end{table}

The evaluation includes three horizons ($h \in \{4, 12, 24\}$ weeks), with the last horizon's observations constituting the test set. The main measure used is MASE~\cite{ref29}, which scales the mean absolute forecast error by the in-sample one-step naive error.
\begin{equation}
\mathrm{MASE} = \frac{\dfrac{1}{h} \displaystyle\sum_{k=1}^{h} |y_{T+k} - \hat{y}_{T+k}|}{\dfrac{1}{T-1} \displaystyle\sum_{t=2}^{T} |y_t - y_{t-1}|}.
\end{equation}
MASE is such a scale-free measure that it is defined whenever the training series is not constant. Also, it is stable as the series approaches zero, where percentage errors are not~\cite{ref29}; it beats naive benchmarks if below one. This avoids having to exclude near-zero series, as in the studies using percentage errors. The sMAPE is reported, which is the symmetric mean absolute percentage error. All metrics are calculated on the original scale after inversion.

\subsection*{Replication, mediation, and selectors}

A single realization is an unreliable basis for inference, so we replicate the experiment over $R$ independent realizations per DGP in replication-major order, with each completed pass yielding a balanced grid; we use $R = 13$, for 35,099 scored forecasts. Comparisons are made within replication. For each transformation, we pair its MASE with the MASE of the untransformed configuration over matching (process, replication, horizon, model) cells and test the paired differences using the Wilcoxon signed-rank test~\cite{ref33}. The resulting $p$-values are corrected using the Benjamini--Hochberg procedure~\cite{ref34}.

The mediation analysis distinguishes (i) whether transformations increase stationarity from (ii) whether stationarity improves accuracy. We evaluate stationarity using three ratios in $[0,1]$, computed on the transformed training series using StationarityToolkit~\cite{ref35}. The ratios are: a trend ratio based on ADF~\cite{ref7} and KPSS~\cite{ref9}; a variance ratio based on the Breusch--Pagan~\cite{ref36} and ARCH~\cite{ref37} tests; and a seasonal ratio based on the autocorrelation at the seasonal lag. Let $T \in \{0,1\}$ indicate whether a transformation is applied, let $S$ denote a stationarity ratio, and let $Y$ denote MASE. For each transformation family, we estimate three relations following Baron and Kenny~\cite{ref38}.
\begin{equation}
S = \alpha_0 + \alpha_1 T,
\end{equation}
\begin{equation}
Y = \beta_0 + \beta_1 S + \beta_2 T,
\end{equation}
\begin{equation}
Y = \gamma_0 + \gamma_1 T.
\end{equation}
These equations define the three pathways described above. Eq~(5) is Path A, the effect of the transformation on stationarity. Eq~(6) is Path B, the effect of stationarity on accuracy while controlling for the transformation. Eq~(7) is Path C, the total effect of the transformation on accuracy. We estimate these relations by transformation family rather than pooling all transformations into a single indicator, which would obscure family-specific mechanisms.

We assess the transformation decision as a choice among five strategies. For each (process, replication, horizon), the oracle is the transformation with the lowest mean MASE across models, and regret is the gap between a strategy's choice and the oracle. The strategies are: no transformation, blanket differencing, blanket detrending, a unit-root-test rule (difference when ADF fails to reject and KPSS rejects; detrend when trend-stationarity is indicated; otherwise none), and an out-of-sample validation selector that chooses the transformation minimizing MASE on the shortest held-out horizon, consistent with the validity of out-of-sample evaluation for autoregressive series~\cite{ref39}.

\section*{Results}

13 Monte Carlo replications of the factorial design yield 35,099 forecast evaluations (of the $2{,}700 \times 13 = 35{,}100$ possible, one cell failed to estimate and is recorded as missing). This section presents aggregate effects, effects by process class, matched versus mismatched transforms, horizon effects, and selector comparisons.

\subsection*{Aggregate effect of transformations}

Table~\ref{table3} presents the mean MASE obtained from each transformation across all the processes, models, and horizons together with the untransformed one (mean MASE 1.984) over matched cells and tested by the Wilcoxon signed-rank test under Benjamini--Hochberg correction. Three transformations enhance the baseline, and these all preserve the conditional-mean signal: seasonal-trend detrending (1.764), linear detrending (1.811), and seasonal differencing (1.841). Any change involving either integer or fractional differencing of the level, or a non-discriminating variance transform, increases MASE, with first differencing at 2.065 and the log-plus-difference pipeline worst at 2.346. All differences are significant after correction, except for seasonal-trend detrending.

\begin{table}[!ht]
\centering
\caption{\textbf{Mean MASE of each transformation, paired against the untransformed configuration (mean MASE 1.984).} Differences are tested by the Wilcoxon signed-rank test with Benjamini--Hochberg correction across the nine comparisons. SE is the standard error of the mean paired difference.}
\begin{tabular}{l l l l l l}
\hline
\textbf{Transformation} & \textbf{Mean MASE} & \textbf{Mean diff.} & \textbf{SE} & $\boldsymbol{p_{BH}}$ & \textbf{Effect} \\
\thickhline
stl\_detrend & 1.764 & $-0.220$ & 0.023 & 0.068 & improves \\
\hline
linear\_detrend & 1.811 & $-0.173$ & 0.020 & $<0.001$ & improves \\
\hline
seasonal\_difference & 1.841 & $-0.143$ & 0.045 & $<0.001$ & improves \\
\hline
fractional\_difference & 2.061 & $+0.077$ & 0.012 & $<0.001$ & worsens \\
\hline
difference & 2.065 & $+0.081$ & 0.032 & $<0.001$ & worsens \\
\hline
log+seasonal\_difference & 2.216 & $+0.232$ & 0.047 & $<0.001$ & worsens \\
\hline
log & 2.226 & $+0.242$ & 0.017 & $<0.001$ & worsens \\
\hline
boxcox & 2.227 & $+0.242$ & 0.020 & $<0.001$ & worsens \\
\hline
log+difference & 2.346 & $+0.362$ & 0.035 & $<0.001$ & worsens \\
\hline
\end{tabular}
\label{table3}
\end{table}

The pattern is that signal-preserving transforms help and signal-removing transforms do not, contrary to the expectation that differencing-based preprocessing is generally beneficial. Because the aggregate pools different process types, it understates the benefit of transforms that are correct for a specific structure; the following subsection resolves this by process class.

\subsection*{Effect by process class}

Table~\ref{table4} reports the mean MASE by process class and principal transformation; the pattern within each row follows the structure of the data-generating process.

\begin{table}[!ht]
\centering
\caption{\textbf{Mean MASE by process class and transformation.} Bold marks the lowest value in each row. Lower is better.}
\begin{adjustwidth}{-1in}{0in}
\centering
\begin{tabular}{l l l l l l l l l}
\hline
\textbf{Class} & \textbf{none} & \textbf{diff.} & \textbf{seas. diff.} & \textbf{frac. diff.} & \textbf{lin. detrend} & \textbf{STL detrend} & \textbf{log} & \textbf{Box--Cox} \\
\thickhline
stationary & 0.88 & 1.15 & 1.28 & 0.89 & 0.88 & 0.88 & \textbf{0.87} & 0.88 \\
\hline
trend\_stationary & 2.79 & 2.08 & 1.02 & 2.87 & 1.95 & \textbf{1.87} & 3.54 & 3.22 \\
\hline
difference\_stationary & 2.46 & 2.51 & 3.75 & 2.72 & \textbf{2.31} & \textbf{2.31} & 2.61 & 2.93 \\
\hline
near\_unit\_root & 2.25 & 2.45 & 3.36 & 2.31 & \textbf{2.24} & 2.25 & 2.25 & 2.25 \\
\hline
fractional & \textbf{0.98} & 1.22 & 1.38 & 1.00 & \textbf{0.98} & 0.99 & \textbf{0.98} & \textbf{0.98} \\
\hline
seasonal\_unit\_root & 3.02 & 4.06 & \textbf{0.44} & 3.13 & 3.20 & 3.04 & 3.13 & 3.14 \\
\hline
seasonal\_stationary & 3.32 & 4.34 & \textbf{0.94} & 3.54 & 3.58 & 3.39 & 3.45 & 3.35 \\
\hline
structural\_break & 1.56 & 1.17 & 1.28 & 1.57 & 1.24 & \textbf{1.09} & 2.34 & 2.55 \\
\hline
heteroscedastic & 1.45 & 1.76 & 1.91 & 1.46 & \textbf{1.44} & 1.45 & 1.46 & 1.45 \\
\hline
heteroscedastic\_trend & 1.04 & 1.32 & 1.32 & 1.07 & \textbf{0.99} & 1.01 & 1.08 & 1.06 \\
\hline
\end{tabular}
\end{adjustwidth}
\label{table4}
\end{table}

\begin{figure}[!h]
\centering
\includegraphics[width=0.85\textwidth]{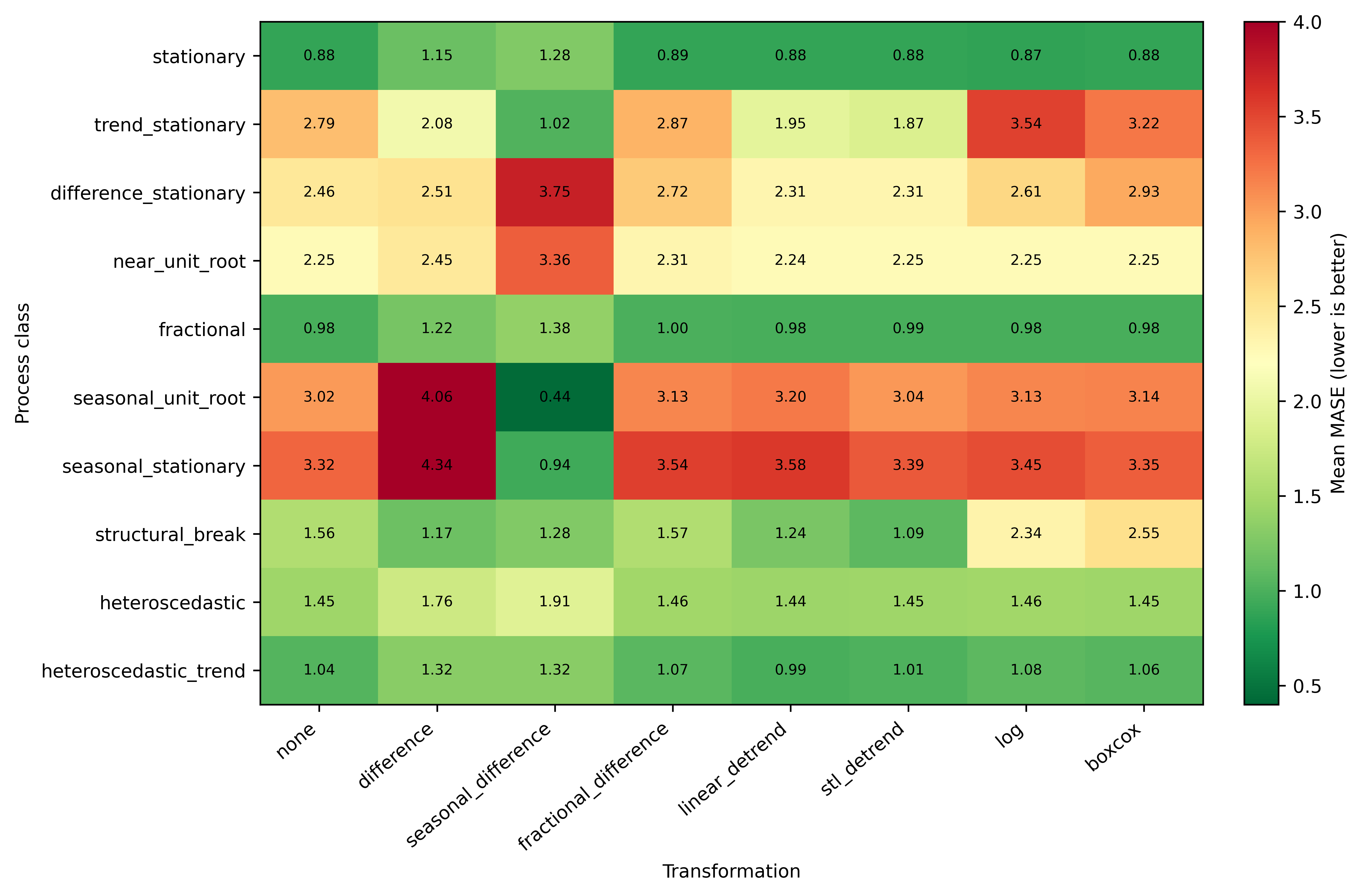}
\caption{\textbf{Mean MASE by process class (rows) and transformation (columns).} Green indicates lower error, and red indicates higher error. The matched transform for each class is the strongest or near-strongest cell in its row.}
\label{fig1}
\end{figure}

On trend-stationary data, detrending and seasonal differencing achieve the lowest error, while the variance-targeting log and Box-Cox transformations are the worst transformations. On the difference-stationary and near-unit-root classes, no transformation improves substantially on no preprocessing, and first differencing is not best, because the models accommodate a stochastic trend internally. On heteroscedastic processes, the variance-stabilizing transforms are best or tied, though the margin is small. The complete matrix is visible in Fig~\ref{fig1}.

\subsection*{Matched transforms and mismatched transforms}

A transform is said to be matched to a process if it targets the known non-stationarity of the process (Table~\ref{table1}). It is said to be mismatched otherwise. According to Table~\ref{table5}, the matching proves to be advantageous. Matched transforms reach a mean MASE of 1.725, whereas the mismatched transforms have a mean MASE of 2.105. Also, they improve on no preprocessing in 53.8 per cent of the cells against 34.6 per cent. The gain from a transformation is thus contingent on its correspondence to the data-generating process and not a generic property of transformation.

\begin{table}[!ht]
\centering
\caption{\textbf{Matched versus mismatched transform-process pairs.} ``Beats none'' is the percentage of cells in which the transform achieves a lower MASE than the untransformed configuration.}
\begin{tabular}{l l l l}
\hline
\textbf{Pair type} & \textbf{n} & \textbf{Mean MASE} & \textbf{Beats none} \\
\thickhline
Matched & 4,680 & 1.725 & 53.8\% \\
\hline
Mismatched & 30,419 & 2.105 & 34.6\% \\
\hline
\end{tabular}
\label{table5}
\end{table}

\subsection*{Horizon effects}

Table~\ref{table6} presents the mean MASE by forecast horizon for four configuration options: the untransformed family, detrending family, first differencing, and seasonal differencing. Across the forecasting horizons, the ranking among strategies is stable, although the differences widen at longer horizons.

\begin{table}[!ht]
\centering
\caption{\textbf{Mean MASE by forecast horizon and transformation family.}}
\begin{tabular}{l l l l l}
\hline
\textbf{Horizon (weeks)} & \textbf{No transform} & \textbf{Detrend} & \textbf{First difference} & \textbf{Seasonal difference} \\
\thickhline
4 & 1.203 & 1.116 & 1.253 & 1.455 \\
\hline
12 & 1.785 & 1.569 & 1.865 & 1.737 \\
\hline
24 & 2.965 & 2.678 & 3.077 & 2.333 \\
\hline
\end{tabular}
\label{table6}
\end{table}

\subsection*{Comparison between preprocessing selectors}

\begin{figure}[!h]
\centering
\includegraphics[width=0.85\textwidth]{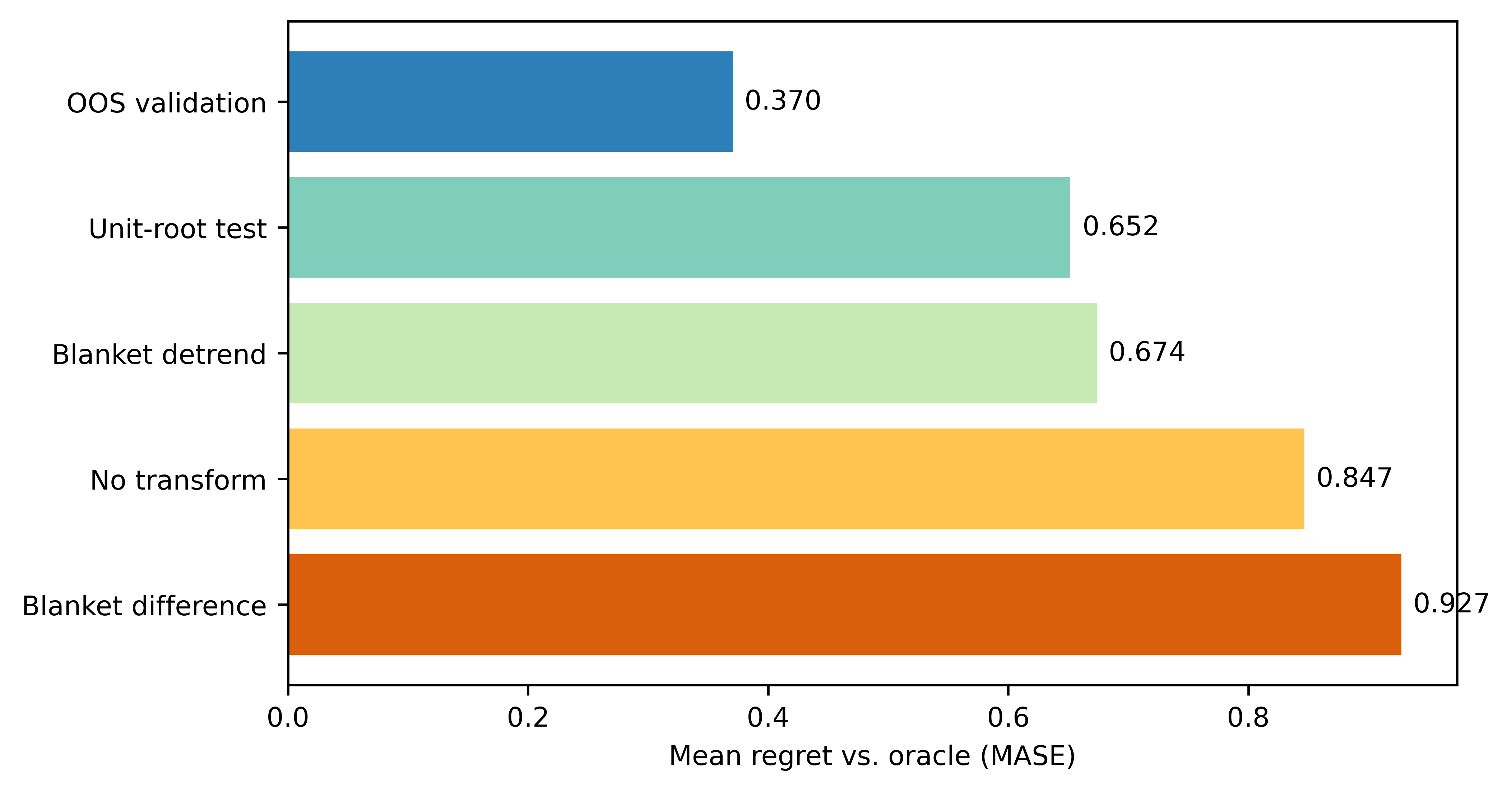}
\caption{\textbf{Mean regret of each preprocessing selector relative to the per-series oracle transformation, in MASE units.} The out-of-sample validation selector attains the lowest regret, followed by the unit-root-test rule; blanket differencing is the worst.}
\label{fig2}
\end{figure}

Table~\ref{table7} shows the mean MASE of the chosen transformation and the mean regret relative to the per-series oracle for the five strategies. The out-of-sample validation selector achieves the lowest regret (0.370), which is 43 percent below that of the unit-root-test rule and 60 percent below blanket differencing. The unit-root-test rule outperforms both fixed differencing and no transformation, consistent with Diebold and Kilian~\cite{ref25}, but it does not match selection by direct out-of-sample evaluation. Blanket differencing has high regret in a differencing-to-stationarity pipeline. Overall, it is better to evaluate candidate transformations on held-out data than to rely on a fixed rule or a stationarity test. Regret is shown in Fig~\ref{fig2}.

\begin{table}[!ht]
\centering
\caption{\textbf{Preprocessing selectors ranked by mean regret compared with the per-series oracle transformation.} Lower regret is better.}
\begin{tabular}{l l l}
\hline
\textbf{Selector} & \textbf{Mean MASE} & \textbf{Mean regret} \\
\thickhline
Out-of-sample validation & 1.508 & 0.370 \\
\hline
Unit-root-test rule & 1.789 & 0.652 \\
\hline
Blanket detrend & 1.811 & 0.674 \\
\hline
No transform & 1.984 & 0.847 \\
\hline
Blanket difference & 2.065 & 0.927 \\
\hline
\end{tabular}
\label{table7}
\end{table}

\section*{Mediation analysis}

The results above show what transformations do to accuracy; here we ask which dimension of stationarity that effect tracks, corroborating the by-class pattern of Table~\ref{table4}. The conventional pipeline assumes that a transformation achieves stationarity and, in turn, the stationarity leads to better forecasts. This section decomposes the chain by the mediation framework, examining the three paths separately for each transform family.

\subsection*{Path coefficients}

Table~\ref{table8} reveals the coefficients. Path A examines whether stationarity improvement occurs along each dimension due to the transformation; Path B analyzes whether lower MASE is attributable to stationarity after transformation under control; Path C estimates the aggregate effect of the transformation on MASE.

\begin{table}[!ht]
\centering
\caption{\textbf{Mediation path coefficients by transform family.} Path A: transformation to stationarity. Path B: stationarity to MASE, controlling for transformation. A positive Path A coefficient indicates the transformation raises that stationarity dimension; a negative Path B coefficient indicates greater stationarity is associated with lower error. Path C: total transformation effect on MASE. Negative Path C indicates the family lowers error on average.}
\begin{adjustwidth}{-1in}{0in}
\centering
\begin{tabular}{l l l l l l l l}
\hline
\textbf{Family} & \textbf{A (trend)} & \textbf{A (var.)} & \textbf{A (seas.)} & \textbf{B (trend)} & \textbf{B (var.)} & \textbf{B (seas.)} & \textbf{C (total)} \\
\thickhline
Differencing & $+0.157$ & $+0.104$ & $+0.028$ & $-0.048$ & $-0.583$ & $-1.653$ & $+0.122$ \\
\hline
Detrending & $+0.177$ & $+0.037$ & $-0.001$ & $-0.062$ & $-0.851$ & $-2.662$ & $-0.196$ \\
\hline
Variance & $-0.009$ & $-0.055$ & $-0.000$ & $-0.375$ & $-1.172$ & $-2.405$ & $+0.242$ \\
\hline
Seasonal & $+0.013$ & $+0.075$ & $-0.065$ & $-0.535$ & $-1.162$ & $-1.672$ & $+0.045$ \\
\hline
\end{tabular}
\end{adjustwidth}
\label{table8}
\end{table}

\subsection*{Interpretation}

Three findings jointly elucidate the process-class results. First, differencing and detrending both attain trend stationarity: their Path A trend coefficients are positive and comparable ($+0.157$ and $+0.177$), confirming both succeed at the statistical task the pipeline sets. As anticipated, the trend ratio is unaffected by the variance and seasonal families.

Second, achieving trend stationarity is not very accurate. For every family, the Path B trend coefficient is small ($-0.048$ to $-0.062$ for the differencing and detrending families that act on the trend), whereas the variance coefficients ($-0.583$ to $-1.172$) and seasonal coefficients ($-1.653$ to $-2.662$) are one order of magnitude larger and negative throughout. In these estimates, accuracy is associated with variance and seasonal stationarity far more than with trend stationarity. This dissociation between the property that the pipeline optimizes and the more desirable properties helps reduce error.

\begin{figure}[!h]
\centering
\includegraphics[width=\textwidth,keepaspectratio]{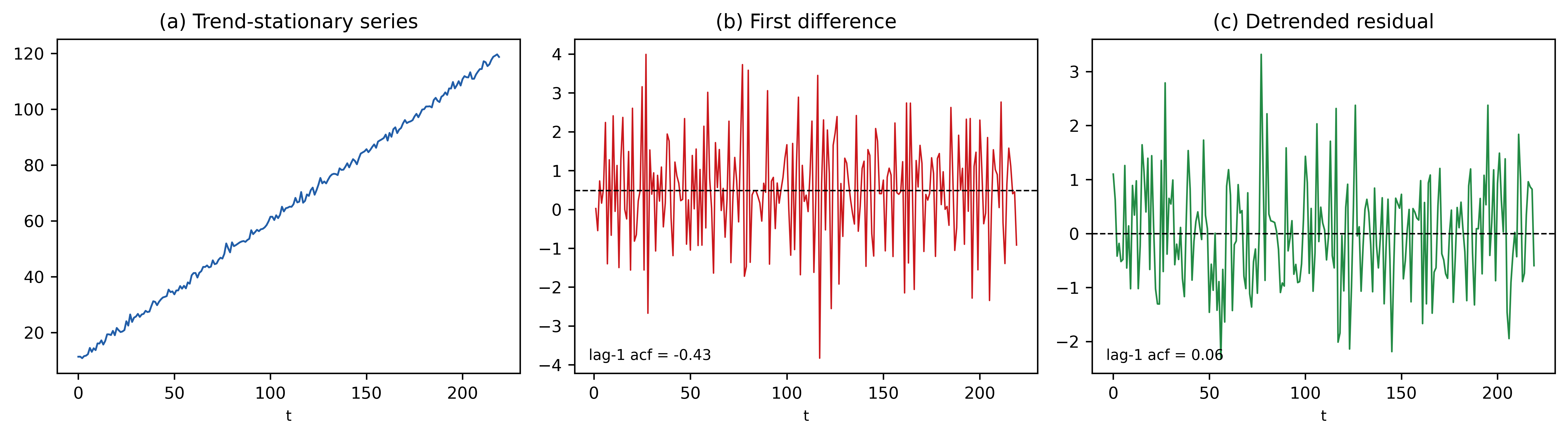}
\caption{\textbf{The over-differencing mechanism on a trend-stationary series.} (a) The original series is dominated by a deterministic linear trend. (b) The first difference removes the trend but leaves a strongly negative lag-one autocorrelation of $-0.43$, the signature of the non-invertible moving-average component induced by over-differencing. (c) The detrended residual, with near-zero autocorrelation of $0.06$, retains the stationary structure that the model can exploit.}
\label{fig3}
\end{figure}

\begin{figure}[!h]
\centering
\includegraphics[width=\textwidth,keepaspectratio]{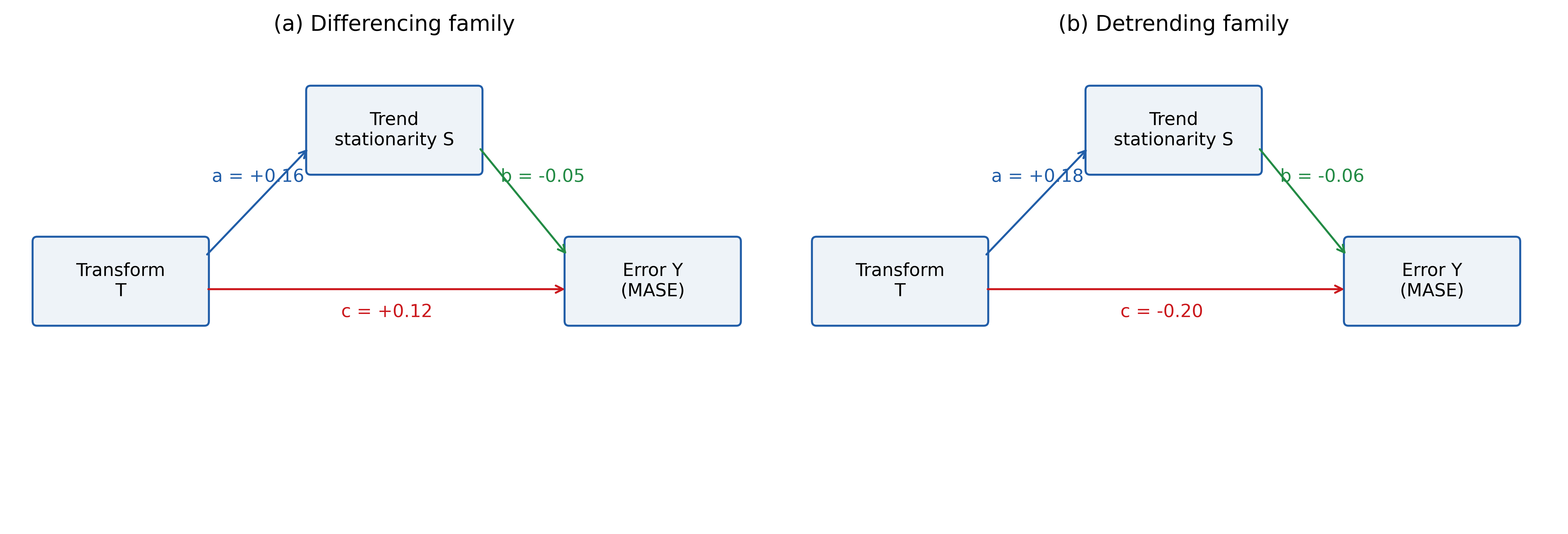}
\caption{\textbf{Mediation path coefficients for the differencing and detrending families on the trend-stationarity dimension.} Both achieve trend stationarity (path \textit{a} positive and comparable), achieved trend stationarity has only a weak association with accuracy (path \textit{b} small for both), and the total effects (path \textit{c}) diverge: detrending lowers error ($c = -0.20$) while differencing raises it ($c = +0.12$).}
\label{fig4}
\end{figure}

Third, total effects diverge for families achieving the same trend stationarity. Only detrending has a negative total effect ($C = -0.196$), so it lowers error on average. Differencing has a comparable effect on trend stationarity but has a positive total effect ($C = +0.122$). The total effects of the variance and seasonal families are $+0.242$ and $+0.045$ when pooled. Their effectiveness is diminished by their influence on processes that do not fall within their scope. The divergence is illustrated at the level of the transformed series in Fig~\ref{fig3}, while the path structure is summarized in Fig~\ref{fig4}.

The difference between the two transforms is not whether they yield stationarity, which they do, but what predictable structure remains. When a deterministic trend is differenced, it works like a high-pass filter. It removes from the series the low-frequency component, which we can best extrapolate as it is most predictable. It also induces a moving-average root not far from the unit circle. The negative lag-one autocorrelation, visible in Fig~\ref{fig3}b, suggests that after differencing this trend, we get a series that is closer to white noise. Detrending estimates and removes the trend so that the deterministic component gets added back to the forecast while the stationary residual retains its short-range dependence. The accuracy ranking mechanism works in this way: signal-preserving transforms eliminate non-stationarity exactly as they do not remove forecastable structure.

The connection between transformation and trend stationarity is present. However, the association between trend stationarity and accuracy is weak. As a result, achieving trend stationarity through differencing does not enhance forecasting accuracy. The transformation decision should be guided by forecastability, even though stationarity as a statistical property and forecastability as a predictive property are related yet distinct.

\section*{Real-world validation}

To test if the synthetic findings hold on data whose true process is unknown, this origin-anchored pipeline was applied to nine real series from two domains: one transportation series (TSA) of United States airport security checkpoint passenger volumes, and eight retail series (Walmart) of weekly unit sales of food products from the M5 forecasting competition dataset. The transportation series displays a strong post-pandemic recovery trend, marked annual seasonality, and changing variance; the retail series are noisier and have a weaker deterministic structure. Together they span the trend-dominated and noise-dominated regimes of the synthetic design. Each series is processed exactly as in the synthetic experiment, giving 1,332 evaluations (of the $9 \times 10 \times 5 \times 3 = 1{,}350$ possible, 18 cells were excluded for numerically unstable inverses): nine series, ten transformations, five models, and three horizons.

\subsection*{Aggregate and domain-specific results}

Table~\ref{table9} presents the mean MASE of each transformation pooled across the eight Walmart retail series, all models, and all horizons. The result is consistent with the synthetic results: no preprocessing and linear detrending are the two best configurations (1.176 and 1.199), while differencing-based transforms reduce accuracy (first differencing 1.459, log-plus-difference 1.754).

\begin{table}[!ht]
\centering
\caption{\textbf{Mean MASE of each transformation on the nine real series, pooled across models and horizons.} Lower is better.}
\begin{tabular}{l l l}
\hline
\textbf{Transformation} & \textbf{Mean MASE} & \textbf{Mean sMAPE} \\
\thickhline
none & 1.176 & 0.248 \\
\hline
linear\_detrend & 1.199 & 0.251 \\
\hline
stl\_detrend & 1.300 & 0.284 \\
\hline
boxcox & 1.302 & 0.320 \\
\hline
fractional\_difference & 1.333 & 0.252 \\
\hline
difference & 1.459 & 0.386 \\
\hline
log & 1.620 & 0.428 \\
\hline
log+difference & 1.754 & 0.490 \\
\hline
seasonal\_difference & 2.017 & 0.558 \\
\hline
\end{tabular}
\label{table9}
\end{table}

\begin{figure}[!h]
\centering
\includegraphics[width=\textwidth,keepaspectratio]{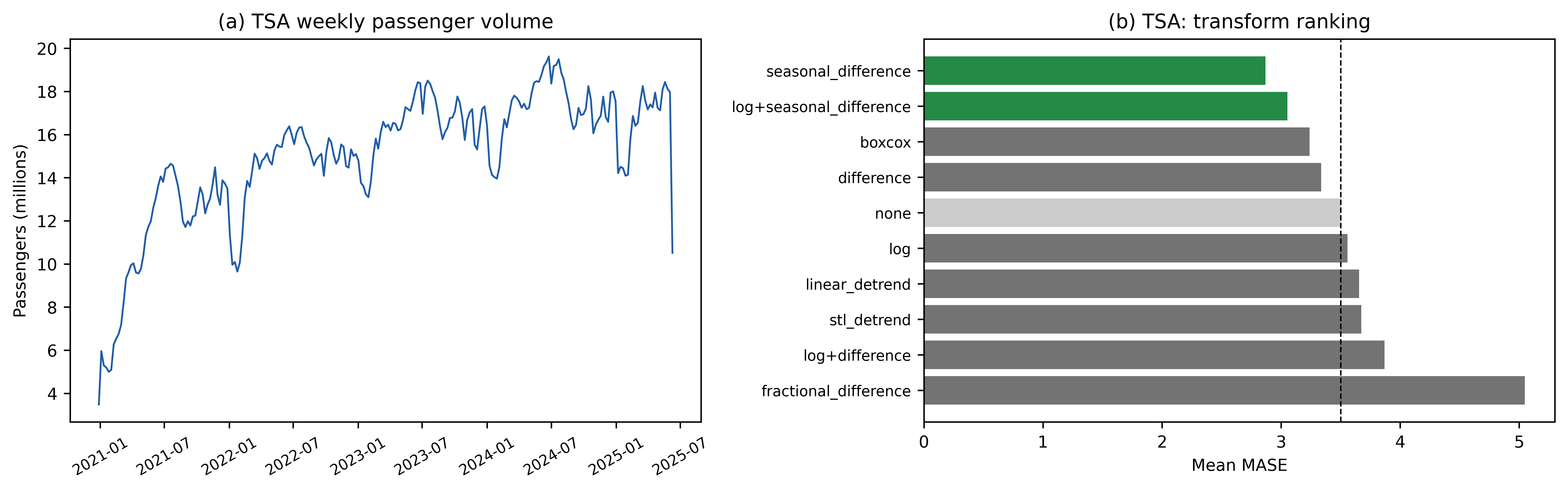}
\caption{\textbf{Real-world validation on the transportation series.} (a) Weekly passenger volume, showing the post-pandemic recovery trend and annual seasonality. (b) Mean MASE by transformation, with the no-preprocessing baseline marked by the dashed line. Seasonal differencing, matched to the series's dominant seasonality, achieves the lowest error.}
\label{fig5}
\end{figure}

Domain resolution enables us to recover the matched-transform pattern, which pooling obscures. For the transportation series, which possesses strong annual seasonality, the best approach is seasonal differencing, which reduces MASE from 3.501 under no preprocessing to 2.869, an eighteen percent reduction. This produces the same matched-transform effect that seasonal differencing produces on the synthetic seasonal classes, but now on seasonality not imposed by construction. For the retail series, which have weak seasonality and weak trend, fractional differencing at 0.868 is marginally better than no preprocessing at 0.886. The reason is consistent with our synthetic finding that fractional differencing preserves the signal in noise-dominated processes. Table~\ref{table10} provides rankings for the transportation series, and the series itself, along with its ranking, is displayed in Fig~\ref{fig5}.

\begin{table}[!ht]
\centering
\caption{\textbf{Mean MASE by transformation on the transportation series (TSA passenger volume), pooled across models and horizons.}}
\begin{tabular}{l l l}
\hline
\textbf{Transformation} & \textbf{Mean MASE} & \textbf{Mean sMAPE} \\
\thickhline
seasonal\_difference & 2.869 & 0.094 \\
\hline
log+seasonal\_difference & 3.053 & 0.099 \\
\hline
boxcox & 3.238 & 0.106 \\
\hline
difference & 3.337 & 0.109 \\
\hline
none & 3.501 & 0.113 \\
\hline
log & 3.558 & 0.115 \\
\hline
linear\_detrend & 3.655 & 0.118 \\
\hline
stl\_detrend & 3.674 & 0.118 \\
\hline
log+difference & 3.869 & 0.123 \\
\hline
fractional\_difference & 5.048 & 0.149 \\
\hline
\end{tabular}
\label{table10}
\end{table}

\subsection*{Selector performance on real series}

On the synthetic grid, the validation choice is averaged over many replications, which stabilizes selection; for a single real series, the validation window is one short held-out segment, so the transform that minimizes error on it can overfit and generalizes imperfectly. In other words, if sufficient held-out data are available to estimate the out-of-sample error reliably, it can be used to select a model. If this is not guaranteed, the gain in out-of-sample performance is not guaranteed over a conservative default on a short single series. The ranking versus blanket differencing, the worst strategy in both settings, survives both settings. Consequently, the real-world outcomes mirror the central synthetic findings under the conditions the synthetic design assumes, and they clarify where those findings transfer and where they do not.

\section*{Discussion}

\subsection*{Principal findings}

Transformations that preserve the conditional-mean signal, efficient detrending, and seasonal differencing applied to the series with the corresponding structure improve accuracy; differencing applied without regard to process type degrades it. The benefit is conditional on the match between the transform and the data-generating process, not an intrinsic property of preprocessing. The mediation analysis seeks to identify the mechanism: differencing and detrending both achieve trend stationarity, yet trend stationarity has a low correlation with accuracy, and the transforms diverge in their effect on the predictable structure that remains. The best way to resolve the preprocessing decision is via out-of-sample evaluation. This out-of-sample evaluation achieves lower regret than unit-root pretesting and every fixed rule. The boundary condition here is that the held-out window needs to be sufficiently large to reliably estimate the out-of-sample error.

\subsection*{Reconciliation with classical theory}

The results conform to the theory of unit roots and differencing and do not differ. The relationship requires precise statement, because the results might otherwise appear to contradict the Box-Jenkins prescription, which they do not.

According to classical theory, differencing is required only when a process is difference-stationary; trend-stationary processes do not require differencing. Applying the incorrect transformation (for example, differencing when the actual underlying data-generating process is trend-stationary) is a specification error~\cite{ref5,ref14,ref15}. We have demonstrated at the level of the data-generating process that the error is raised by differencing and lowered by detrending on the trend-stationary classes; the over-differencing of Plosser and Schwert~\cite{ref14} and the spurious structure of Nelson and Kang~\cite{ref15}, which we observe here as a measured accuracy penalty and visualized in the induced negative moving-average autocorrelation of Fig~\ref{fig3}. The theory predicts the direction of every effect we observe, and our contribution is to quantify its magnitude across models and horizons and show it is large enough to matter. The over-stationarization observed in deep forecasters~\cite{ref40} is, in our controlled setting, the same mechanism by which indiscriminate differencing removes low-frequency predictable structure from classical models.

In the near-unit-root domain, the trend type remains weakly identified in finite samples~\cite{ref10,ref11}. Thus, it follows that the gain from correctly differencing one that is close to a unit root is minor and doesn't suffice to offset the cost of differencing one that is not.

The connection to Diebold and Kilian~\cite{ref25} is an extension, not a contradiction. The authors showed for the AR(1)-with-trend process that unit-root pretesting does better than always differencing or never differencing, and our results reproduce this ordering: the unit root test does better than blanket differencing, which does better than no transformation. The ordering maintains validity across a heterogeneous pool of processes and five model families, and a strategy outside of their three, selection by direct out-of-sample validation, achieves lower regret than pretesting when there is enough held-out data. The result bolsters the decision-theoretic perspective on preprocessing while changing the proposed instrument from a stationarity test to an out-of-sample criterion.

\subsection*{Implications for practice}

Three recommendations follow. When the series is long enough to support an out-of-sample evaluation, that evaluation, rather than a stationarity test alone, should be the basis for the transformation choice. Differencing should not be a standard first step: it is correct for a genuine unit-root process, but on the trend-stationary and weakly-structured processes commonplace in practice, it removes predictable structure, and modern models accommodate trends without it. When a model assumption is violated due to the presence of heteroscedasticity, variance-stabilizing transforms should be applied. These address the variance instability without attenuating the conditional-mean signal. When the time series is short, a conservative default of no preprocessing is preferable to both blanket strategies.

\subsection*{Limitations}

The conclusions depend on four limitations. While the synthetic processes encompass deterministic and stochastic trends, fractional integration, seasonal unit roots, structural breaks, and heteroscedasticity, the design is simpler than many real series, which combine these with regime changes and nonlinear dynamics. The real-world validation addresses this partially in two domains. The model set consists of classical, state-space, and gradient-boosted families, but does not include deep-learning forecasters, which may interact differently with preprocessing. The mediation analysis's stationarity ratios rely on tests whose asymptotic justifications are approximate given the sample length we study, and so the mediation coefficients describe association at this sample size rather than asymptotic quantities. The benefit of selecting based on validation is conditioned on sufficient held-out data and does not transfer to a single very short series; we leave characterizing the series length at which this becomes valid to future work. These bounds define the scope without constraining the main conclusion that the efficiency of a stationarity transformation is a function of its match to the process and is better assessed empirically than by stationarity testing.

\section*{Conclusion}

This study considers whether transforming time series data to make them stationary helps with forecasting them into the future. The decision to transform or not is the object of study and not the premise of the inquiry. Across 35,099 forecasts from a balanced factorial design over eighteen data-generating processes, ten transformations, five models, and three horizons, replicated by Monte Carlo and validated on nine real series from two domains, the evidence supports one organizing conclusion: the value of a stationarity transformation depends on whether it matches the data-generating process, and is better assessed by out-of-sample evaluation than by stationarity testing.

This is supported by three outcomes. Transformations that preserve the conditional-mean signal, such as detrending and seasonal differencing matched to structure, improve accuracy, while indiscriminate differencing degrades it. Further, matched transforms outperform mismatched in mean error and in how often they beat no preprocessing. The paper shows that differencing and detrending both induce trend stationarity, but that trend stationarity is weakly associated with accuracy. The two transforms undermine the predictable structure in different ways. Differencing creates a non-invertible moving-average component due to over-differencing. Of the preprocessing strategies, selection by out-of-sample validation attains the lowest regret on the synthetic grid, ahead of unit-root pretesting but well ahead of blanket differencing, the worst strategy in both synthetic and real-world settings.

\section*{Supporting information}

\paragraph*{S1 File.}
\label{S1_File}
\textbf{Complete experimental code and results.} All experimental code, configuration files, generated synthetic data, the full grid of results, and the analysis scripts that reproduce every table and figure in this article are publicly available at the repository listed in the Data Availability Statement.

\section*{Acknowledgments}

The authors thank Prof. Qibo Yang (Ningbo Polytechnic University) for reviewing the manuscript and providing valuable feedback and insights. During the preparation of this manuscript, the authors used generative AI tools for editing and grammar checking. The authors reviewed and edited the content as needed and take full responsibility for the content of the publication.

\nolinenumbers

\end{document}